\documentclass{PoS}
\usepackage{epsfig}

\newcommand{\beq}{\begin{align*}}
\newcommand{\eeq}{\end{align*}}

\title{
\begin{picture}(0,0)(0,0)%
   \put(350,75){\makebox(0,0)[l]{\textnormal{\normalsize CHIBA-EP-161}}}%
\end{picture}%
Quark confinement and gauge invariant monopoles in SU(2) YM}

\ShortTitle{Quark confinement and gauge invariant monopoles in SU(2) YM}

\author{
   S.~Kato$^a$\thanks{Speaker at the conference(kato@takamatsu-nct.ac.jp).}, S.Ito$^b$, 
   K.-I.~Kondo$^c$,
   T.~Murakami$^d$, A.~Shibata$^e$ 
   and T.~Shinohara$^d$\\
\llap{$^a$}Takamatsu National College of Technology, Takamatu City 761-8058, Japan\\
\llap{$^b$}Nagano National College of Technology, 716 Tokuma, Nagano 381-8550, Japan\\
\llap{$^c$}Department of Physics, Faculty of Science, Chiba University, Chiba 263-8522, Japan\\
\llap{$^d$}Graduate School of Science and Technology, Chiba University, Chiba 263-8522, Japan
\llap{$^e$}Computing Research Center, High Energy Accelerator Research Organization (KEK), Tsukuba 305-0801, Japan
}

\abstract{
 We give a short review of recently obtained results on a new lattice formulation of the 
non-linear change of variables which was once called the Cho--Faddeev--Niemi decomposition
in  SU(2) Yang-Mills theory. 
 Based on this formulation, we proposed a new gauge-invariant definition of the magnetic 
monopole current which guarantees the magnetic charge quantization. 
We also demonstrated the magnetic monopole dominance in the string tension in SU(2) Yang-Mills 
theory on a lattice. 
Our formulation enables one to reproduce in the gauge-invariant way remarkable  results obtained so 
far only in the Maximally Abelian gauge. }

\FullConference{XXIVth International Symposium on Lattice Field Theory\\
                July 23-28, 2006\\
                Tucson, Arizona, USA}

\begin{document}

\section{Introduction}

It is interesting to study color confinement mechanism in Quantum Chromodynamics (QCD).
The dual superconductor scenario of the QCD vacuum may be a candidate for that
mechanism. In paticular, it is known that the string tension calculated from the abelian
and monopole parts reproduces well the original one, once we perform an abelian projection
in Maximally Abelian (MA) gauge. It is so-called "abelian and monopole dominance". But 
it has been diffcult to see these phenomena in any other gauges. One of the reason of this 
gauge problem may come from the fact that they used the monopoles on the lattice defined 
by DeGrand and Toussaint(DT) \cite{DT80}. 

Recently, we have demonstrated that the gauge-invariant magnetic monopole  can be constructed 
in the pure Yang-Mills theory without any fundamental scalar field. 
The success is achieved based on a new viewpoint proposed by three of us \cite{KMS05} for the 
non-linear change of variables (NLCV), which was called Cho--Faddeev--Niemi (CFN) decomposition 
\cite{Cho80}\cite{FN98}, see also \cite{Shabanov99}. 
We have found that the magnetic charge of our lattice magnetic monopole is perfectly quantized.
Moreover, we have confirmed dominance of our magnetic monopole  in the string tension, while it was 
first shown in \cite{SNW94} in the conventional Maximally Abelian (MA) gauge \cite{KLSW87}. 
Therefore we can show the gauge invariance of the dual superconductor scenario of the QCD vacuum.

In this paper, we summarize the recent results on a lattice formulation of Yang-Mills theory 
based on NLCV and a gauge-invariant magnetic monopole on a lattice\cite{IKKMSS06}.

\section{Lattice CFN variables or NLCV on a lattice}

We have proposed a natural and useful lattice formulation of  the non-linear change of 
variables (NLCV) in Yang-Mills theory corresponding to the CFN decomposition \cite{Cho80,FN98}. 
It is a minimum requirement that such a lattice formulation must reproduce  the continuum counterparts 
in the naive continuum limit. 
In this stage, therefore, it is instructive to recall how the CFN variables are defined in the continuum 
formulation. 
We restrict the following argument to SU(2) gauge group, for simplicity.%

In the continuum formulation \cite{Cho80,KMS05}, a color vector field $\vec{n}(x)=(n_A(x))$ $(A=1,2,3)$ 
is introduced as a three-dimensional unit vector field. 
In what follows, we use the boldface to express the Lie-algebra $su(2)$-valued field, e.g., 
${\bf n}(x) :=n_A(x)T_A$, $T_A=\frac{1}{2}\sigma_A$ with Pauli matrices $\sigma_A$ ($A=1,2,3$).
Then  the $su(2)$-valued gluon field (gauge potential) ${\bf A}_\mu(x)$ is decomposed into two parts:
\begin{eqnarray}
  {\bf A}_\mu(x) = {\bf V}_\mu(x) + {\bf X}_\mu(x) ,
\end{eqnarray}
in such a way that the color vector field ${\bf n}(x)$ is covariantly constant in the background  
field ${\bf V}_\mu(x)$:
\begin{eqnarray}
 0 = {\cal D}_\mu[{\bf V}] {\bf n}(x) 
:= \partial_\mu {\bf n}(x) -i g [{\bf V}_\mu(x) , {\bf n}(x) ],
 \label{covariant-const}
\end{eqnarray}
and that the remaining field ${\bf X}_\mu(x)$ is perpendicular to ${\bf n}(x)$:
\begin{eqnarray}
  \vec{n}(x) \cdot \vec{X}_\mu(x) \equiv 2{\rm tr}({\bf n}(x)   {\bf X}_\mu(x)) = 0 .
  \label{nX0}
\end{eqnarray}
Here we have adopted the normalization ${\rm tr}(T_A T_B)= \frac12 \delta_{AB}$. Both 
${\bf n}(x)$ and ${\bf A}_\mu(x)$ are Hermitian fields. This is also the case for $ {\bf V}_\mu(x)$ 
and ${\bf X}_\mu(x)$. 
By solving the defining equation (\ref{covariant-const}), the   ${\bf V}_\mu(x)$ field is obtained in the form:
\begin{eqnarray}
  {\bf V}_\mu(x) 
  = c_\mu(x) {\bf n}(x)   -i g^{-1} [ \partial_\mu {\bf n}(x), {\bf n}(x) ] .
  \label{Vdef}
\end{eqnarray}

On a lattice, on the other hand, we introduce the site variable ${\bf n}_{x}$ , 
in addition to the original link variable $U_{x,\mu}$ which is related to the gauge 
potential ${\bf A}_\mu(x)$ in a naive way: 
\footnote{
In general, the argument of the exponential in (\ref{def-U}) is the line integral of a gauge potential 
along a link from $x$ to $x+\mu$.
 Note also that we define a color vector field 
${\bf n}(x) :=n_A(x)T_A$ in the continuum, while ${\bf n}_x  :=n_x^A\sigma_A$ on the lattice for convenience.

}
\begin{eqnarray}
U_{x,\mu} = \exp( -i \epsilon g {\bf A}_\mu(x)) , 
\label{def-U}
\end{eqnarray}
where $\epsilon$ is the lattice spacing and  $g$ is the coupling constant. Here  ${\bf n}_{x}$ is 
Hermitian, ${\bf n}_{x}^\dagger={\bf n}_{x}$, and $U_{x,\mu}$ is unitary, $U_{x,\mu}^\dagger=U_{x,\mu}^{-1}$.   
The link variable $U_{x,\mu}$ and the site variable ${\bf n}_{x}$ transform under the gauge 
transformation II \cite{KMS05} as
\begin{eqnarray}
  U_{x,\mu} \rightarrow \Omega_{x} U_{x,\mu} \Omega_{x+\mu}^\dagger = U_{x,\mu}' , \quad
  {\bf n}_{x} \rightarrow \Omega_{x} {\bf n}_{x} \Omega_{x}^\dagger = {\bf n}_{x}' .
\end{eqnarray}

Suppose we have obtained a link variable $V_{x,\mu}$ as a group element of $G=SU(2)$, which is 
related to the  $su(2)$-valued background field ${\bf V}_\mu(x)$ through 
\begin{eqnarray}
  V_{x,\mu} = \exp (-i\epsilon g {\bf V}_\mu(x)) ,
\end{eqnarray}
where ${\bf V}_\mu(x)$ is to be identified with the continuum variable (\ref{Vdef}) and hence 
 $V_{x,\mu}$ must be unitary $V_{x,\mu}^\dagger=V_{x,\mu}^{-1}$.

A lattice version of (\ref{covariant-const}) and (\ref{nX0}) is respectively given by
\begin{eqnarray}
 {\bf n}_{x} V_{x,\mu}  = V_{x,\mu} {\bf n}_{x+\mu} ,
 \label{Lcc}
\end{eqnarray}
and
\begin{equation}
 {\rm tr}({\bf n}_{x} U_{x,\mu} V_{x,\mu}^\dagger) 
  = 0 .
  \label{cond2m}
\end{equation}
 Both conditions must be imposed to determine $V_{x,\mu}$ for a given set of ${\bf n}_{x}$ and 
$U_{x,\mu}$. 

By solving the defining equation (\ref{Lcc}) and (\ref{cond2m}),
the link variable $V_{x,\mu}$ is obtained up to an overall normalization constant 
in terms of the site variable ${\bf n}_{x}$ and the original link variable 
$U_{x,\mu} = \exp( -i \epsilon g {\bf A}_\mu(x))$, 
just as the continuum CFN variable and 
${\bf A}_\mu(x)$ in (\ref{Vdef})\footnote{See \cite{IKKMSS06} for the detail.}:
\begin{eqnarray}
  V_{x,\mu} = V_{x,\mu}[U,{\bf n}] 
  = U_{x,\mu} +  {\bf n}_{x} U_{x,\mu} {\bf n}_{x+\mu} .
  \label{sol}
\end{eqnarray}

Finally, the unitary link variable $\hat{V}_{x,\mu}[U,{\bf n}]$ is obtained after the normalization: 
\begin{eqnarray} 
\hat{V}_{x,\mu} = 
\hat{V}_{x,\mu}[U,{\bf n}] := 
 V_{x,\mu}/\sqrt{\frac{1}{2}{\rm tr} [V_{x,\mu}^{\dagger}V_{x,\mu}]} .
\label{cfn-mono-4}
\end{eqnarray}
It is easy to show that the naive continuum limit $\epsilon \rightarrow 0$ 
of the link variable (\ref{cfn-mono-4}) reduces to the continuum expression (\ref{Vdef}). 

Therefore, we can define the {\it gauge-invariant flux}, $\bar{\Theta}_P[U,{\bf n}]$,
(plaquette variable) by
\begin{eqnarray} 
\bar{\Theta}_{x,\mu\nu}[U,{\bf n}] := \epsilon^{-2}
{\rm arg} ( {\rm tr} \{({\bf 1}+{\bf n}_x)\hat{V}_{x,\mu}\hat{V}_{x+\hat{\mu},\nu}
\hat{V}_{x+\nu,\mu}^{\dagger}\hat{V}_{x,\nu}^{\dagger} \}/{\rm tr}({\bf 1})) .
\label{cfn-mono-5}
\end{eqnarray}
It is also shown that the naive continuum limit of (\ref{cfn-mono-5}) reduces to
the gauge-invariant field strength;
\begin{eqnarray} 
\bar{\Theta}_{x,\mu\nu} \simeq
\partial_{\mu}c_{\nu}-\partial_{\nu}c_{\mu}
+ g^{-1} {\bf n}\cdot (\partial_{\mu}{\bf n}\times\partial_{\nu}{\bf n})
\equiv G_{\mu\nu}(x) ,
\label{cfn-fs} 
\end{eqnarray}
which plays the similar role that 'tHooft--Polyakov tensor played in describing the 
magnetic monopole in Gergi--Glashow model.

\section{Numerical simulations}

It has been shown that the SU(2) master Yang-Mills theory written in terms of 
${\bf A}_\mu(x)$ and ${\bf n}(x)$ has the enlarged local gauge 
symmetry $\tilde{G}^{\omega,\theta}_{local}=SU(2)_{local}^{\omega} \times [SU(2)/U(1)]_{local}^{\theta}$ 
larger than the local gauge symmetry $SU(2)_{local}^{\omega}:=SU(2)_{local}^{I}$ in the original 
Yang--Mills theory. 
  In order to fix the whole enlarged local gauge symmetry $\tilde{G}^{\omega,\theta}_{local}$, 
we must impose sufficient number of gauge fixing conditions. 

First of all, we generate the configurations of SU(2) link variables 
$\{ U_{x,\mu} \}$, $ U_{x,\mu}= \exp [ - ig\epsilon {\bf A}_\mu(x) ]$,
using the standard Wilson action based on the heat bath method. 
Next, we define the new Maximal Abelian gauge  (nMAG) on a lattice.  
By introducing a vector field ${\bf n}_{x}$ of a unit length with three components, 
we consider a functional $F_{nMAG}[U, {\bf n}; \Omega, \Theta]$  written in terms 
of  the gauge (link) variable $U_{x,\mu}$ and the color (site) variable ${\bf n}_{x}$ 
defined by 
\begin{eqnarray}
F_{nMAG}[U, {\bf n}; \Omega, \Theta]
\equiv \sum_{x,\mu}  {\rm tr}({\bf 1}-{}^{\Theta}{\bf n}_{x} 
{}^{\Omega}U_{x,\mu} {}^{\Theta}{\bf n}_{x+\mu} {}^{\Omega}U_{x,\mu}^\dagger ) .
\end{eqnarray} 
Here we have introduced the enlarged gauge transformation: 
$
 {}^\Omega{}U_{x,\mu} := \Omega_{x} U_{x,\mu} \Omega_{x+\mu}^\dagger
$
for the  link variable $U_{x,\mu}$ and
$
 {}^{\Theta}{\bf n}_{x} := \Theta_{x} {\bf n}_{x}^{(0)} \Theta_{x}^\dagger  
$ 
for an initial site variable ${\bf n}_{x}^{(0)}$
where gauge group elements $\Omega_{x}$ and $\Theta_{x}$   are independent SU(2) matrices on a site $x$.
The former corresponds to the  $SU(2)^{\omega}$ gauge transformation $({\bf A}_\mu)^{\omega}(x)$  of the 
original potential: 
$
({\bf A}_\mu)^{\omega}(x)=\Omega(x)[{\bf A}_\mu(x)+ig^{-1} \partial_\mu]\Omega^\dagger(x)
={\bf A}_\mu(x)+D_\mu[{\bf A}]\bf{\omega}(x) + O({\bf \omega}^2)$ for the 
$\Omega_{x}=e^{i g{\bf \omega}(x)}$, 
while 
the infinitesimal form of the latter reads   
$
{\bf n}^{\theta}(x)
  ={\bf n}(x)+g{\bf n}(x) \times {\bf \theta}(x) 
  ={\bf n}(x)+g{\bf n}(x) \times {\bf \theta}(x) 
$
for the adjoint $[SU(2)/U(1)]^{\theta}$ rotation
$\Theta_{x}=e^{i g{\bf \theta}(x)}$. 

After imposing the nMAG, the theory still has the local gauge symmetry 
$SU(2)_{local}^{\omega=\theta}:=SU(2)_{local}^{II}$, since the 'diagonal' 
gauge transformation ${\bf \omega}={\bf \theta}$ does not change the vaue 
of the functional $F_{nMAG}[U, {\bf n}; \Omega, \Theta]$. Therefore, 
${\bf n}_{x}$ configuration can not be determined at this stage. 
In order to completely fix the gauge and to determine ${\bf n}_{x}$, 
we need to impose another gauge fixing condition for fixing  $SU(2)_{local}^{II}$.  
For example, we choose the  conventional Lorentz-Landau gauge or 
Lattice Landau gauge (LLG) for this purpose. 
The LLG can be imposed by minimizing the function 
$F_{LLG}[U; \Omega]$:  
\begin{eqnarray}
  F_{LLG}[U; \Omega] = \sum_{x,\mu} {\rm tr}({\bf 1}-{}^{\Omega}U_{x,\mu}) 
  \rightarrow 1/4 \int d^4x \  [({\bf A}_\mu)^{\omega}(x)]^2  
  \quad (\epsilon \rightarrow 0) ,
\end{eqnarray}
with respect to the gauge transformation $\Omega_{x}$ for the given link 
configurations $\{ U_{x,\mu} \}$.

 In what follows, the desired color vector field ${\bf n}_{x}$ is constructed  
from the interpolating gauge transformation matrix $\Theta_{x}$ by choosing the initial 
value  ${\bf n}_{x}^{(0)}=\sigma_3$ and 
\begin{eqnarray}
  {\bf n}_{x} := \Theta_{x} \sigma_3 \Theta_{x}^\dagger
  = n_{x}^A \sigma^A  , 
  \quad   n_{x}^A =  {\rm tr}[\sigma_A \Theta_{x} \sigma_3 \Theta_{x}^\dagger]/2   \quad  (A=1,2,3).
\end{eqnarray}

We generate the configurations of the color vector field $\{{\bf n}_x\}$ according to the method 
explained above together with the configurations of SU(2) link variables $\{U_{x,\mu}\}$. 
Then we can construct $\{\hat{V}_{x,\mu}[U,{\bf n}]\}$ from (\ref{cfn-mono-4}).
The numerical simulations are performed on an $8^4$ lattice   at $\beta=$2.2, 2.3, 2.35, 2.4, 
2.45, 2.5, 2.6 
and on $16^4$ lattice at $\beta=2.4$ by thermalizing 3000 sweeps respectively. 

\subsection{Quantization of magnetic charges}

We construct the gauge-invariant field strength (\ref{cfn-mono-5}) to extract  configurations of 
the magnetic monopole current $\{k_{x,\mu}\}$  defined by
\begin{eqnarray}
k_{\mu}(s)=
-\frac{1}{4\pi}{\varepsilon}_{\mu\nu\rho\sigma}
\partial_{\nu}\bar{\Theta}_{\rho\sigma}(x+\mu)
\simeq
-\frac{1}{4\pi}{\varepsilon}_{\mu\nu\rho\sigma}
\partial_{\nu}G_{\rho\sigma}(x) .
\label{cfn-conti-20}
\end{eqnarray}
This definition agrees with our definition of the monopole in the continuum (divided by $2\pi$).

In our formulation, we have only the real variable $\bar{\Theta}_P[U,{\bf n}]$ 
at hand, and we are to calculate the monopole current using the final term in (\ref{cfn-conti-20}).  
Therefore, it is not so trivial to obtain the integer-valued $k_{\mu}(s)$ from the real-valued 
$\bar{\Theta}_P[U,{\bf n}]$.  
To check quantization of the magnetic charge, we have made a histogram of
$
 K(s,\mu) := 2\pi k_\mu(s) = \frac{1}{2}{\varepsilon}_{\mu\nu\rho\sigma}
\partial_{\nu}\bar{\Theta}_{\rho\sigma}(x+\mu)
$,
 i.e., magnetic charge distribution. 
 Note that $K(s,\mu)$ 
  should become a multiple of $2\pi$ if the magnetic charge is quantized.  
Our numerical results show that $K(s,\mu)$ is completely separated into $0$ or $\pm 2\pi$ within an 
error of $10^{-10}$, see 
Table~\ref{table:hist-of-rmg}. 
We have checked that the data in  Table~\ref{table:hist-of-rmg}  exhaust in total all the 
configurations $N=4\times 8^4=16384$,
because the number $N_l$ of links in the $d$-dimensional lattice with a side length $L$ is 
given by  $N_l=dL^d$. 
This result clearly shows that the new CFN magnetic charge is quantized as expected from the 
general argument. 
We have observed that the conservation law of the monopole current holds, since the number of 
$+2\pi$ configurations is the same as that of $-2\pi$ configurations.

\begin{table}[h]
\caption{Histogram  of the magnetic charge (value of $K(s,\mu)$)   distribution 
for CFN monopoles on  $8^4$ lattice at $\beta=2.35$. }
\label{table:hist-of-rmg}
\begin{center}
\begin{tiny}
\begin{tabular}{c|c|c|c|c|c|c|c|c|c|c|c|c|c|c|c}\hline
Charge & -7.5 & -6.5 & -5.5 & -4.5 & -3.5 & -2.5 & -1.5 & -0.5 
   .    & 0.5 & 1.5 & 2.5 & 3.5 & 4.5 & 5.5 & 6.5 \\
       &$\sim$-6.5 & $\sim$-5.5 & $\sim$-4.5 & $\sim$-3.5 & $\sim$-2.5 & $\sim$-1.5 & $\sim$-0.5 
& $\sim$0.5 & $\sim$1.5 & $\sim$2.5 & $\sim$3.5 & $\sim$4.5 & $\sim$5.5 & $\sim$6.5 & $\sim$7.5 \\ \hline
Number & 0 & 299 & 0 & 0 & 0 & 0 & 0 & 15786 & 0 & 0 & 0 & 0 & 0 & 299 & 0 \\ \hline
\end{tabular}
\end{tiny}
\end{center}
\end{table}
\vspace{-0.7cm}

\subsection{Monopole dominance of the string tension}

In order to study the monopole dominance in the string tension, we proceed to estimate the 
magnetic monopole contribution $\left< W_m(C) \right>$ to the Wilson loop average, i.e., 
the expectation value of the Wilson loop operator $\left< W_f(C) \right>$.  
We define the magnetic part $W_m(C)$ of the Wilson loop operator $W_f(C)$ as the contribution 
from the monopole current $k_{\mu}(s)$ to the Wilson loop operator:
\footnote{
The Wilson loop operator $W_f(C)$ is decomposed into the magnetic part $W_m(C)$ and the electric 
part $W_e(C)$, which is derived from the non-Abelian Stokes theorem, see Appendix B of \cite{Kondo00}.
In this paper, we do not calculate the electric contribution $\left< W_e(C) \right>$ where 
$W_e(C)$ is expressed by the electric current $j_{\mu}=\partial_{\nu}F_{\mu\nu}$. 
}
\begin{eqnarray}
 W_m(C)     &=& \exp(2\pi i \sum_{s,\mu}k_{\mu}(s)N_{\mu}(s)) ,
 \label{monopole dominance-1}     
\\
N_{\mu}(s)  &=& \sum_{s'}\Delta_L^{-1}(s-s')\frac{1}{2}
\epsilon_{\mu\alpha\beta\gamma}\partial_{\alpha}
S^J_{\beta\gamma}(s'+\hat{\mu}), 
\quad
\partial'_{\beta}S^J_{\beta\gamma}(s) = J_{\gamma}(s) ,
\label{monopole dominance-2}
\end{eqnarray}
where $N_{\mu}(s)$ is defined through the external electric source $J_{\mu}(s)$ which is used 
to calculate the static potential:
$\partial'$ denotes the backward lattice derivative
$\partial_{\mu}^{'}f(x)=f(x)-f(x-\mu)$,  $S^J_{\beta\gamma}(s)$ denotes a surface  with the 
closed loop $C$ on which the electric source $J_{\mu}(s)$ has its support, and $\Delta_L^{-1}(s-s')$ 
is the Lattice Coulomb propagator. 
We obtain the string tension by evaluating the average of (\ref{monopole dominance-1}) from the 
generated configurations of the monopoles $\{k_{\mu}(s)\}$.

We calculate the respective potential $V_i(R)$ from the respective average $\left<W_i(C)\right>$:
\begin{eqnarray}
V_i(R) = -\log \left\{ \left< W_i(R,T) \right>/\left<W_i(R,T-1)\right> \right\} 
\quad (i=f, m)
\label{monopole dominance-3}
\end{eqnarray}
where $C=(R,T)$ denotes the Wilson loop $C$ with side lengths $R$ and $T$.  The obtained 
numerical potential is fitted to a linear term,  Coulomb term and a constant term: 
\begin{eqnarray}
V_i(R) = \sigma_i R -  \alpha_i/R +c_i ,
\label{monopole dominance-4}
\end{eqnarray}
where $\sigma$ is the string tension, $\alpha$ is the Coulomb coefficient, 
and $c$ is the coefficient of the perimeter decay:
$\left<W_i(R,T)\right> \sim \exp [-\sigma_i RT -c_i(R+T)+\alpha_i T/R + \cdots]$. 
The numerical simulations are performed on $8^4$ lattice at $\beta=2.3, 2.4$, and  $16^4$ 
lattice at $\beta=2.4$, 50 configurations in each case.  
The results are shown in 
 Table~\ref{strint-tension-1}. 
In order to obtain the full SU(2) results $\sigma_f$, $\alpha_f$, especially, we used the 
smearing method \cite{albanese87} as noise reduction techniques for 100 configurations 
on $16^4$ lattice.

\begin{table}[h]
\caption{String tension and Coulomb coefficient I}
\label{strint-tension-1}
\begin{center}
\begin{tabular}{cllll}\hline
$\beta$ & $\sigma_f$ & $\alpha_f$ & $\sigma_m$ & $\alpha_m$ \\ \hline
2.3($8^4$)  & 0.158(14)   & 0.226(44)  &  0.135(13) & 0.009(36)  \\
2.4($8^4$)  & 0.065(13)   & 0.267(33)  &  0.040(12) & 0.030(34)  \\
{\bf 2.4($16^4$)} & 0.075(9)   & 0.23(2)  &  {\bf 0.068(2)}  & 0.001(5)\\
\hline
\end{tabular}
\end{center}
\end{table}
\vspace{-.5cm}

\begin{table}[h]
\caption{String tension and Coulomb coefficient II (reproduced from \cite{SNW94})}
\label{strint-tension-2}
\begin{center}
\small
\begin{tabular}{cllll}\hline
$\beta$ & $\sigma_f$ & $\alpha_f$ & $\sigma_{DTm}$ & $\alpha_{DTm}$ \\ \hline
{\bf 2.4($16^4$)}  & 0.072(3)   & 0.28(2)  &  {\bf 0.068(2)} & 0.01(1)  \\
2.45($16^4$) & 0.049(1)   & 0.29(1)  &  0.051(1) & 0.02(1)  \\
2.5($16^4$)  & 0.033(2)   & 0.29(1)  &  0.034(1) & 0.01(1) \\
\hline
\end{tabular}
\end{center}
\end{table}
\vspace{-0.5cm}

We find that the numerical errors at $\beta=2.3$ of $8^4$ lattice are relatively small in 
spite of  small size of the lattice $8^4$. Moreover, the monopole part $\sigma_m$ reproduces 
85$\%$ of the full string tension $\sigma_f$. 
The data of   $8^4$ lattice at $\beta=2.4$ show large errors. 
The data of $16^4$ lattice at $\beta=2.4$ exhibit relatively small errors for the full potential
and shows that $\sigma_m$ reproduces 91$\%$ of $\sigma_f$. 
In general, the monopole part does not include  the Coulomb term and hence the potential 
is obtained to an accuracy better than the full potential. 

For comparison, we have shown in Table~\ref{strint-tension-2} the data of \cite{SNW94} 
which has discovered the monopole dominance for the first time on $16^4$ lattice where 
$\sigma_{DTm}$ reproduces 95$\%$ of $\sigma_f$. 
Here $\sigma_{DTm}$ and $\alpha_{DTm}$ denotes the conventional monopole contribution 
extracted from the diagonal potential $A_\mu^3$ using  Abelian projection in MAG. 
In particular, the comparison between the data on $16^4$ lattice at $\beta=2.4$ and the 
data  on $16^4$ lattice at $\beta=2.4$ in Table~\ref{strint-tension-1} reveals that the 
monopole contributions have the same value between the conventional DT monopole and our
magnetic monopole.
Thus, we have confirmed the monopole dominance in the string tension using our magnetic
monopole.

\section{Conclusion and discussion}

In this paper, we have proposed a new formulation of the NLCV of Yang-Mills theory which was 
once called the CFN decomposition.  
This compact formulation of new variables enables us to guarantee the magnetic charge quantization.
The monopole dominance has been shown anew in the string tension. 
Thus, the magnetic charge quantization and monopole dominance in the string tension are confirmed 
in the gauge invariant way, whereas they have been so far shown only in a special gauge fixing 
called  MA gauge which breaks the color symmetry explicitly. 

\section{Acknowledgments}

The numerical simulations have been done on a supercomputer (NEC SX-5) at Research Center for 
Nuclear Physics (RCNP), Osaka University. This work is also supported by the Large Scale Simulation 
Program No.06-17 (FY2006) of High Energy Accelerator Research Organization (KEK). 
This work is in part supported by Grant-in-Aid for Scientific Research (C) 18540251  from Japan Society 
for the Promotion of Science (JSPS), and by Grant-in-Aid for Scientific Research on Priority Areas (B)13135203 
from the Ministry of Education, Culture, Sports, Science and Technology (MEXT).

\end{document}